# Kinetics of phase separation, border of miscibility gap in Fe-Cr and limit of Cr solubility in iron at 832 K

S. M. Dubiel[1*] and J. Żukrowski[2]

[1]AGH University of Science and Technology, Faculty of Physics and Applied Computer Science, al. A. Mickiewicza 30, 30-059 Kraków, Poland, [2]AGH University of Science and Technology, Academic Center for Materials and Nanotechnology, al. A. Mickiewicza 30, 30-059 Kraków, Poland

**Abstract**

Kinetics of phase decomposition accompanied by precipitation of σ-phase in a $Fe_{73.7}Cr_{26.3}$ alloy isothermally annealed at 832 K was studied by means of Mössbauer spectroscopy. Two stage decomposition process has been revealed by three different quantities viz. the average hyperfine field, <H>, the short-range parameter, $\alpha_1$, and the probability of atomic configuration with no Cr atoms within the first two coordination shells around the probe Fe atoms, P(0,0). The first stage, that has terminated after ~300 h of annealing, has been associated with the decomposition into Fe-rich phase in which the concentration of Cr, determined as 20.9 at.%, can be interpreted as the border of the metastable miscibility gap at 832 K. The second stage can be regarded as a continuation of the phase decomposition process combined with a precipitation of σ. The three relevant parameters for this stage have also saturation-like behavior vs. annealing time and the saturation can be interpreted as termination of the two processes. The concentration of Cr in the Fe-rich phase has been determined as 19.8 at.% and this value can be regarded as the limit of Cr solubility in iron at 832 K. Both stages of the kinetics were found to be in line with the Johnson-Mehl-Avrami-Kolgomorov equation yielding values of the rate constant and the Avrami exponent. The activation energy of the second-stage process was determined to be by ~12 kJ/mol higher.

*Corresponding author: Stanislaw.Dubiel@fis.agh.edu.pl

# 1. Introduction

Fe-Cr alloys belong to the most frequently investigated binary alloys. This stems, on one hand, form their interesting magnetic properties and on the other hand from their usefulness in the steel making industry. Concerning the latter the Fe-Cr alloys constitute the major ingredient of a family of stainless steels, in which ferritic/martensitic (FM) ones play an important role as structural materials used in various branches of industry. This role follows from their excellent properties like good resistance to a high-temperature corrosion, low swelling and high toughness. Consequently, they have been used in various branches of industry e. g. power pants (including nuclear ones), chemical and petrochemical industries to produce devices that work at service at elevated temperatures and often in aggressive environment. Concerning the nuclear power plants, for example, their life time is limited by a degradation of structural devices like vessel and primary circuit due to exposure to radiation and high temperature. The former causes radiation damage and the latter thermal aging. Both result in degradation of mechanical properties and corrosion. The main reasons for this degradation are precipitation of: (1) Cr-rich $\alpha'$ phase and (2) $\sigma$-phase. Both effects can be connected with the crystallographic phase diagram of the Fe-Cr system. The precipitation of $\alpha'$ occurs at temperatures below ~770 K and it was originally detected by annealing at 748 K (475°C), so it is known as "475°C embrittlement" [1]. The concentration of Cr in $\alpha'$ is higher than ~85 at%, and this is the true reason of the brittleness. The $\sigma$-phase may precipitate if the annealing temperature is in the range of ~770-1100 K, and the content of Cr lies between ~15 and ~85 at.% Cr. The precipitation of $\alpha'$ is a consequence of the so-called phase separation leading to formation of Fe-rich ($\alpha$) and Cr-rich ($\alpha'$) phases. The phase field in which this process takes place is known as a miscibility gap (MG). An interest in this phenomenon is two-fold. On one hand, one wants to know mechanism(s) underlying the segregation and, on the other hand, borders of MG. Based on numerous studies two mechanisms have been proposed: (1) nucleation and growth, and (2) spinodal. The latter is active in the central part of MG, while the former on both "sides" of the spinodal [2]. In other words the phase fields of the nucleation and growth are close to the Fe-rich and Cr-rich borders of MG. From the technological view-point the Fe-rich border is of greater importance because it is located close to the Cr concentration at which the alloy becomes stainless i.e. ~10.5 at% (In fact the border line is temperature dependent). This border line can be also interpreted as the solubility limit of Cr in iron. Consequently, a great body of papers, both theoretical and experimental, were devoted to the issue. Nevertheless, a clear cut picture has not been obtained yet, because the former give different predictions e. g. [2-9], and the latter, as reported in [3], show a wide spread of data obtained with different techniques [3]. On the other hand, the solubility limit values determined with the Mössbauer spectroscopy (MS) exhibit a systematic trend [11-16], so application of this method is hoped to deliver a set of data that can be used for validation of different predictions. In this paper we report such data obtained for a $Fe_{73.7}Cr_{26.3}$ sample isothermally annealed at 832 K for up to 1777 h.

## 2. Experimental

The study was carried out on a ~25 μm thick foil in form a 20x20 mm rectangle obtained by rolling down ~100 μm thick tape of a $Fe_{73.7}Cr_{26.3}$ alloy. The alloy was made by melting in an induction furnace under protective Ar atmosphere proper quantities of Armco-iron and chromium of 99.9% purity. The ingot was next rolled down to the thickness of ~100 μm. Its composition was determined by a chemical analysis. To promote the decomposition process, the sample was isothermally annealed under dynamic vacuum (<$10^{-4}$ Torr) at 832 K for up to 1777 h. After each annealing, a $^{57}$Fe Mössbauer spectrum was recorded at room temperature in a transmission mode using a standard spectrometer with a drive working in a sinusoidal mode. 14.4 keV gamma rays were emitted by a $^{57}$Co/Rh source whose activity permitted to record statistically good spectrum in 1024 channels of a multichannel analyzer within a 2 days run.

Each spectrum was analyzed assuming that an effect of Cr atoms situated in the first-two neighbor shells around $^{57}$Fe probe nuclei, 1NN-2NN, on the hyperfine field, H, and on the center shift, CS, was additive i.e. $X(m,n) = X(0,0) + m\Delta X_1 + n\Delta X_2$, where X=H or CS, $\Delta X_k$ is a change of X due to one Cr atom situated in 1NN (k=1) or in 2NN (k=2). This procedure has already proved to work properly when analyzing spectra recorded on different Fe-based binary alloys including Fe-Cr ones e. g. [12,15,17,18]. The total number of possible atomic configurations (m,n) in the 1NN-2NN approximation amounts to 63. However, for x = 26.3 at% most of them have negligible probabilities, consequently 17 most probable (according to the binomial distribution) were included into the fitting procedure (their overall probability was > 0.97). However, their probabilities, P(m,n), (associated with spectral areas of sextets corresponding to the selected configurations) were considered as free parameters in the fitting routine. Free parameters were also *X(0,0),* line width (common to all sextets), *G, ΔH$_1$, ΔH$_2$*, and an angle between the magnetization vector and normal to the sample's surface, Θ. On the other hand, following our previous studies values of ΔCS$_1$ = -0.02 mm/s, and ΔCS$_1$= -0.01 mm/s, were kept constant [17]. The spectra annealed for 1537 and 1777 hours showed additional low-intensity line in the central part of the corresponding spectra. This sub spectrum has been associated with precipitated σ-phase. In addition, some spectra were analyzed in terms of a magnetic hyperfine field distribution method [20] to better visualize the effect of annealing.

Examples of the spectra and corresponding hyperfine field distribution curves, p(H), are presented in Figs. 1 and 2, respectively. They evidently illustrate a redistribution of Cr atoms that had taken place on the applied annealing. In the spectra the effect is best seen in the outermost lines, and in the p(H)-curves a shift towards a higher value is evident as well as an increase in the intensity of the peak indicating the maximum value of <H>. This peak can be associated with the (0,0) atomic configuration.

Table 1 Best-fit spectral parameters. P(0,0) is in %; H(0,0), ΔH$_1$, ΔH$_2$ and <H> in kOe; <IS>, IS$_σ$, G and G$_σ$ are in mm/s. The meaning of the symbols is given in the text.

| t [h] | P(0,0) | H(0,0) | ΔH$_1$ | ΔH$_2$ | <H> | <IS> | Θ [°] | G | G$_σ$ | IS$_σ$ |
|---|---|---|---|---|---|---|---|---|---|---|
| 0 | 6.1(3) | 334.3(6) | -34.1(3) | -22.0(2) | 263.6 | -0.142 | 45.2 | 0.27 | | |
| 0.12 | 6.3(3) | 334.8(5) | -33.5(3) | -21.8(1) | 266,7 | -0.141 | 57.3 | 0.25 | | |
| 0.5 | 6.6(3) | 336.4(5) | -33.6(2) | -21.6(2) | 266.4 | -0.142 | 57.8 | 0.26 | | |
| 0.83 | 7.1(3) | 333.7(4) | -33.5(2) | -22.0(1) | 266.6 | -0.139 | 57.0 | 0.26 | | |
| 3 | 7.6(4) | 333.2(4) | -33.5(2) | -22.0(2) | 265.5 | -0.140 | 57.5 | 0.25 | | |
| 4 | 7.8(4) | 335.7(6) | -33.7(3) | -21.9(2) | 266.4 | -0.141 | 58.0 | 0.28 | | |
| 7 | 8.3(4) | 335.6(4) | -32.9(2) | -21.7(2) | 266.0 | -0.137 | 59.3 | 0.25 | | |
| 10 | 8.7(4) | 333.6(5) | -33.6(5) | -22.1(2) | 266.3 | -0.138 | 58.3 | 0.25 | | |
| 34 | 8.8(4) | 336.1(3) | -34.4(2) | -22.6(2) | 267.8 | -0.140 | 59.9 | 0.25 | | |
| 45 | 9.2(5) | 334.0(5) | -34.7(3) | -22.4(2) | 267.0 | -0.141 | 58.4 | 0.26 | | |
| 58 | 9.4(3) | 337.6(4) | -32.2(2) | -21.1(1) | 278.5 | -0.137 | 62.9 | 0.24 | | |
| 79 | 9.5(3) | 336.0(3) | -33.1(2) | -21.9(1) | 270.0 | -0.138 | 61.0 | 0.24 | | |
| 100 | 9.5(3) | 336.5(3) | -32.8(2) | -22.1(3) | 270.3 | -0.139 | 61.5 | 0.23 | | |
| 121 | 9.8(3) | 337.1(4) | -33.0(2) | -22.0(2) | 270.6 | -0.140 | 62.4 | 0.23 | | |
| 145 | 10.0(4) | 336.8(4) | -32.7(3) | -21.8(3) | 270.9 | -0.139 | 62.0 | 0.24 | | |
| 168 | 10.0(4) | 336.8(3) | -32.2(2) | -21.9(1) | 271.2 | -0.138 | 61.2 | 0.23 | | |
| 214 | 10.1(2) | 337.0(3) | -32.9(2) | -21.7(1) | 271.3 | -0.137 | 62.4 | 0.24 | | |

| 303 | 10.2(3) | 335.8(4) | -32.2(2) | -21.9(2) | 271.4 | -0.136 | 61.2 | 0.24 | | |
| 398 | 10.4(2) | 335.9(3) | -33.0(2) | -21.9(1) | 272.0 | -0.138 | 60.7 | 0.25 | | |
| 609 | 10.5(3) | 336.8(3) | -32.5(2) | -21.6(1) | 273.3 | -0.138 | 61.7 | 0.25 | | |
| 1014 | 10.9(2) | 337.3(2) | -33.0(2) | -22.0(1) | 274.5 | -0.136 | 62.0 | 0.25 | | |
| 1537 | 12.2(2) | 338.2(2) | -32.0(2) | -21.3(2) | 275.1 | -0.139 | 62.2 | 0.24 | 0.30 | -0.25 |
| 1777 | 12.3(3) | 337.5(2) | -32.6(3) | -21.8(2) | 275.2 | -0.138 | 62.6 | 0.24 | 0.30 | -0.25 |

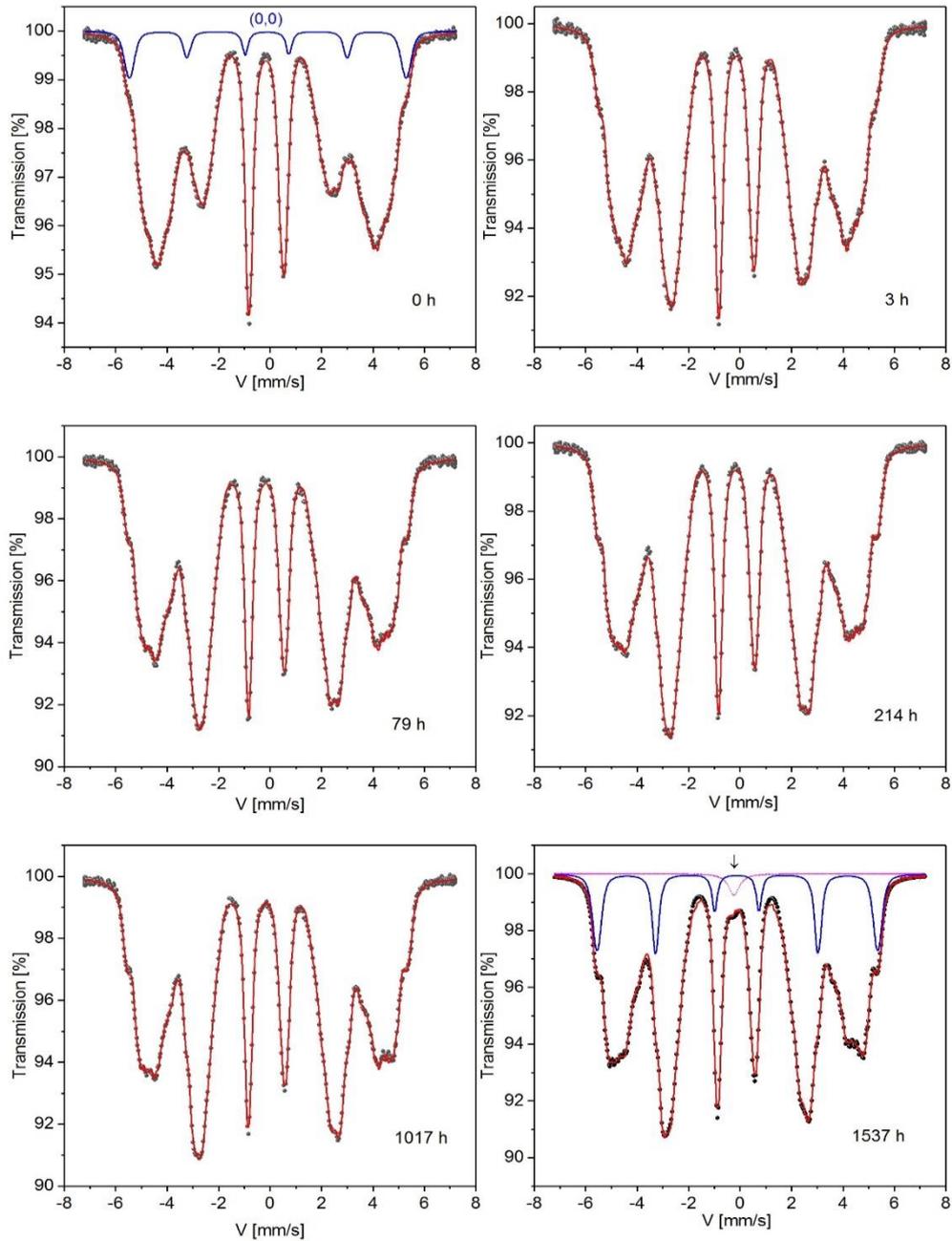

Fig. 1 Selected $^{57}$Fe Mössbauer spectra recorded at 295 K on the studied sample. Each spectrum is labelled with the annealing time. In the untreated spectrum and in the one annealed for 1537 h the component corresponding to the (0,0) atomic configuration is shown. In the latter case the arrow indicates a single-line sub spectrum associated with the σ-phase.

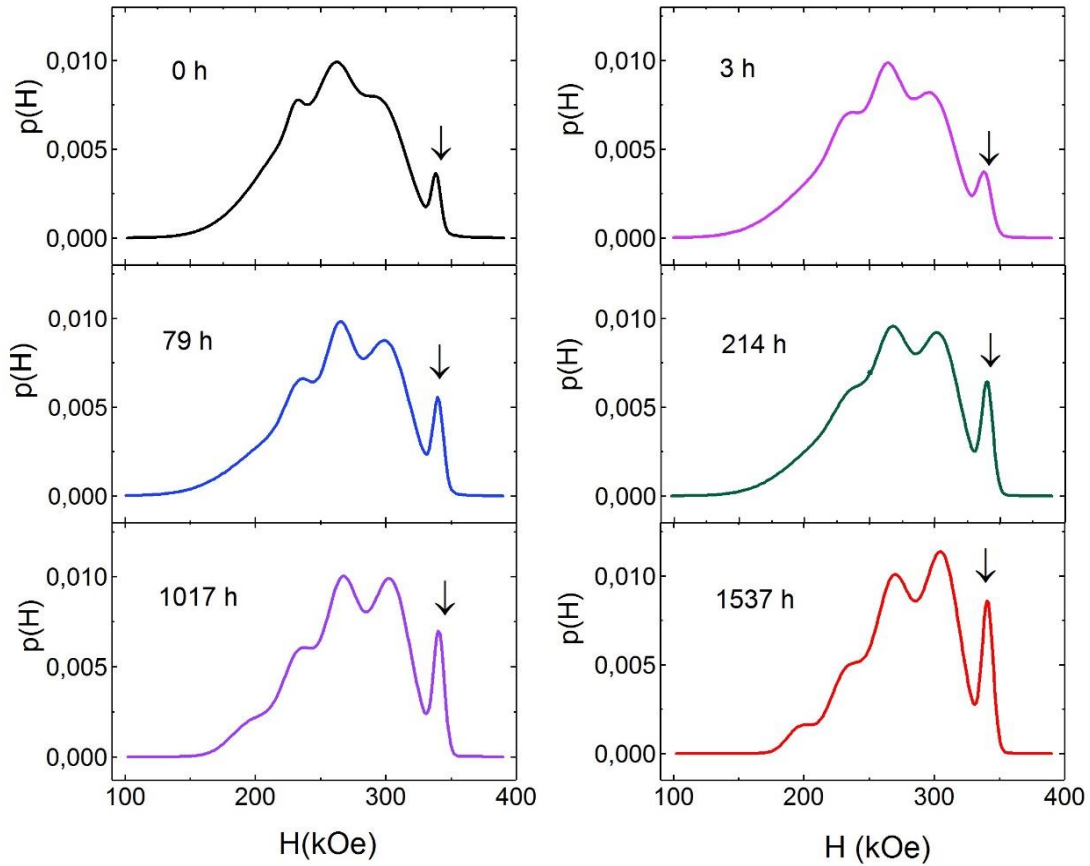

Fig. 2 Distributions of the magnetic hyperfine field derived from the spectra shown in Fig. 1. Notice an increase with the annealing time of the intensity of the peak situated at the maximum value of the field. This peak, marked by arrow, corresponds to the (0,0) atomic configuration.

## 3. Results

### 3.1. Kinetics of phase separation

The kinetics of the phase separation governed by the nucleation and growth can be properly described in terms of the Johnson-Mehl-Avrami-Kolgomorov (JMAK) equation. In our previous studies of this process, also by using the Mössbauer spectroscopy, we have shown that the kinetics can be well reproduced by applying the JMAK law to the average hyperfine field, <H> [15,16,18]:

$$<H> = <H>_o + [1 - \exp(-(kt)^n)] \qquad (1)$$

Where $k$ is the rate constant, $n$ is the Avrami exponent, and $t$ stands for time. The rate constant is related via the Arrhenius law to the activation energy, $E$, as follows:

$$k = k_o e^{-E/k_B T} \qquad (2)$$

Here $k_B$ stays for the Boltzmann constant and $T$ for temperature. Equation (2) can be used to determine the activation energy of a given process, $E$. For this purpose one has to measure its kinetics at two different temperatures, $T_1$ and $T_2$. Then based on Eq. (2) the following equation for $E$, can be derived:

$$E = (T_1 T_2 / (T_2 - T_1)) k_B \ln\left(\frac{k_1}{k_2}\right) \qquad (3)$$

Using this method for $T_1$=681 K and $T_2$=722 K we have found $E$=122 kJ/mol (1.26 eV) for $Fe_{85}Cr_{15}$ [15]. The temperature dependence of <H> in the present case is shown in Fig. 3. It is evident that two processes take place: (I) in the time interval up to ~300 h with the activation energy $E_1$, and (II) for higher $t$-values with the activation energy $E_2$. Applying again Eq.(2) the following relation between $E_1$ and $E_2$ can be found:

$$E_2 = E_1 + k_B T \ln\left(\frac{k_1}{k_2}\right) \qquad (4)$$

It can be rewritten as:

$$E_2 = E_1 + \Delta E \qquad (4a)$$

Where $\Delta E$ is a difference between the two energies. Putting into Eq.(2) values of the rate constants obtained by fitting <H>-data to the JMAK equation yields $\Delta E$=11.6 kJ/mol (0.12 eV). For determining the absolute value of $E_1$, hence that of $E_2$, one would need to know the $k$-values obtained for a given sample at two different temperatures. Recently we performed a study of the phase separation on the identical sample at 800 K getting $k$=0.020(3) h$^{-1}$ [16]. This means that the rate constant has within the error limit the same value as the one at 832 K, hence the Eq. (3) cannot be used. Consequently, we can determine only the difference between $E_2$ and $E_1$.

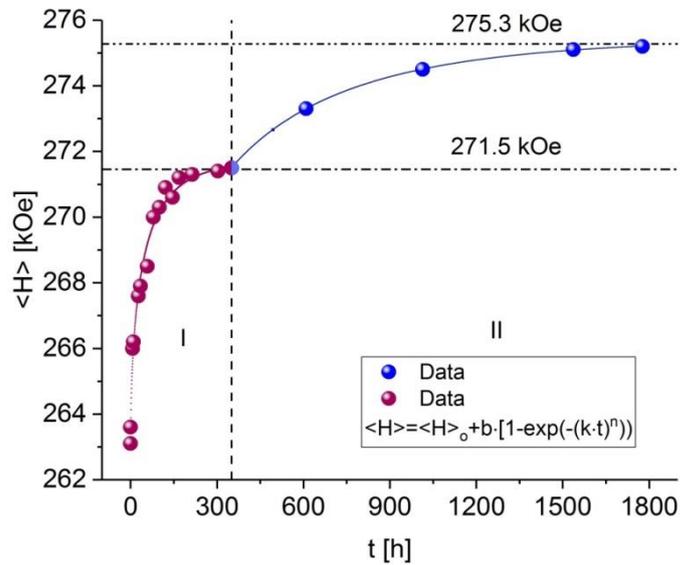

Fig. 3 The average hyperfine field, <H>, vs. annealing time, $t$. The solid lines represent the best-fits to the data in terms of the JMAK equation in range I and II, respectively.

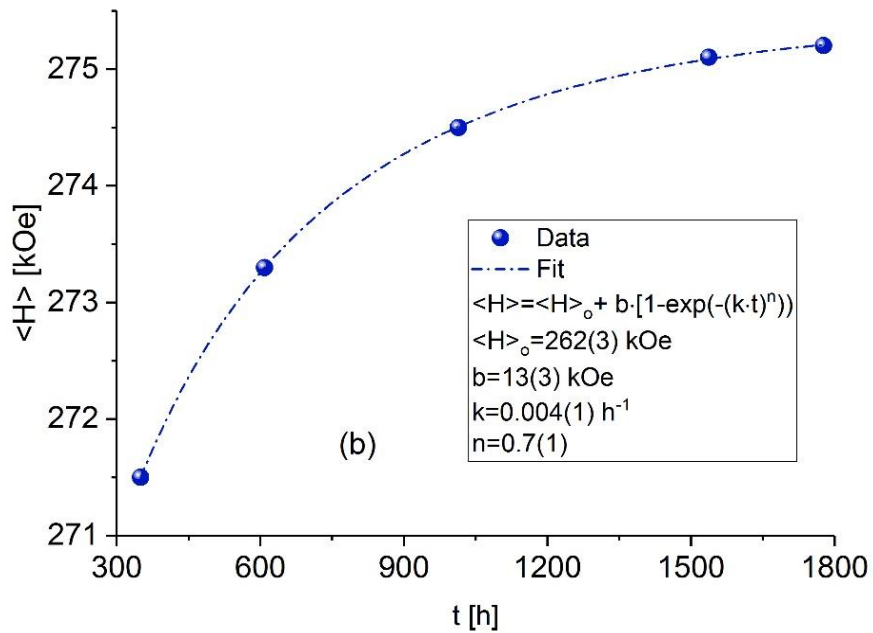

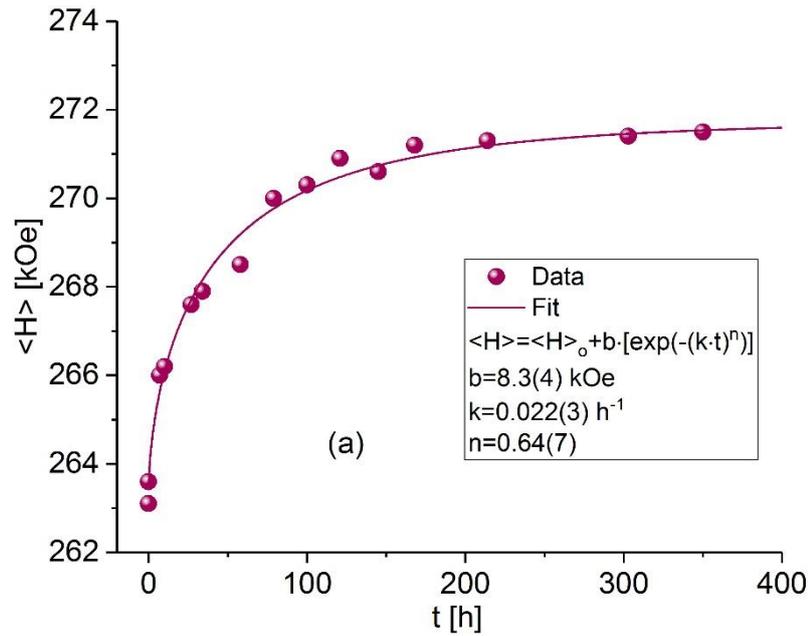

Fig. 4 The average hyperfine field, <H>, vs. annealing time, $t$, for the first (a), and for the second (b) stage of the decomposition. The solid line represents the best-fit to the data in terms of the JMAK equation. The best-fit values of the kinetics parameters i.e. rate constant, $k$, and the Avrami exponent, $n$, are given in the legends.

**3.2. Limit of Cr solubility**

The content of Cr in the Fe-rich phase, $x$, hence the limit of Cr solubility in iron can be determined based on a quasi-linear relationship between <H> and $x$ [12]. For this purpose one takes the value of <H> in saturation, $<H>_s$. In the present case the phase decomposition has two stages: I and II, as indicated in Fig. 2. $<H>_s$=271.5 kOe in the stage I which corresponds to $x_I$=20.7 at.%, and in the stage II $<H>_s$=275.3 kOe which yields $x_{II}$=19.7 at.%. The first stage being metastable can be associated with the metastable border of MG. The Cr concentration determined for the second stage can be regarded as the solubility limit of Cr at 832 K. These values of $x$ can be next used to validate some calculations relevant to the issue. In particular, in Fig. 4 we have made a comparison of our present and previous data with the prediction concerning the Fe-rich branch of MG as reported by Bonny et al. [21], in Fig. 5 with the recent calculations of the full MG by Jacob et al. [10], and in Fig. 6 with the complete phase diagram as reported by Bony et al. [4]. Concerning Fig. 4, a distinction between the two predictions shown in this figure is not possible due to large errors of the potential method. Measurements et higher temperature are needed.

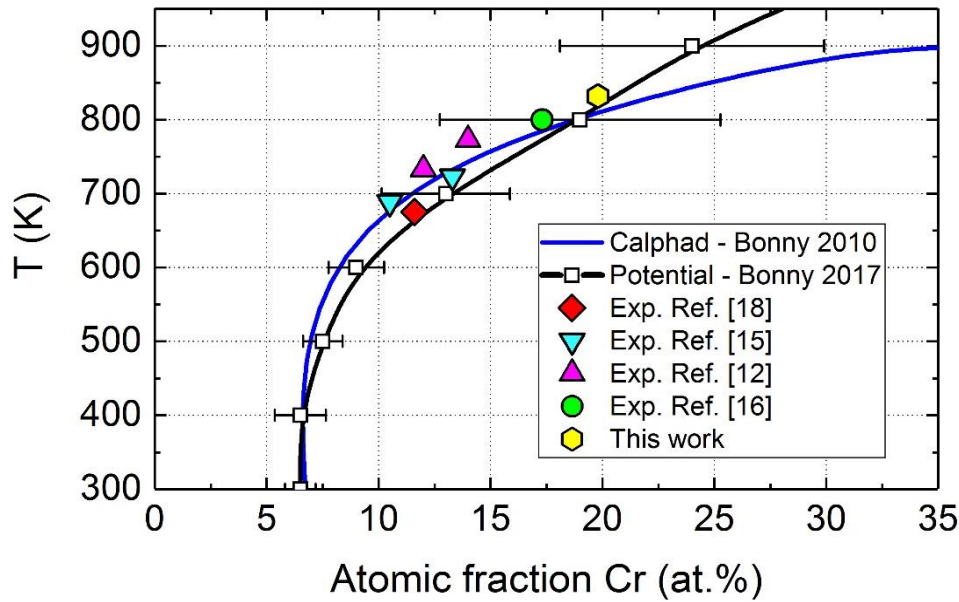

Fig. 5 Comparison between two theoretical predictions concerning the Fe-rich branch of the MG [20] and experimental data obtained with the Mössbauer spectroscopy. The size of the symbols is comparable with the error bar.

In turn the data shown in Fig. 5 fit best to the Calphad prediction due to Xiong et al. [6], at least as far as the Fe-rich MG border is concerned. The data depicting the Cr-rich MG border are in agreement with all three predictions shown in this figure. Finally, concerning the calculated complete phase diagrams of the Fe-Cr system shown in Fig. 7 the experimental data obtained with the Mössbauer spectroscopy agree better with the one predicted by Bonny et al. [4], also the difference between the two predictions is rather minor above ~650 K. The main disagreement concerns the border between the real and metastable miscibility gap. According to our results the border should lie between 800 and 832 K, while the predicted temperature is lower than 800 K.

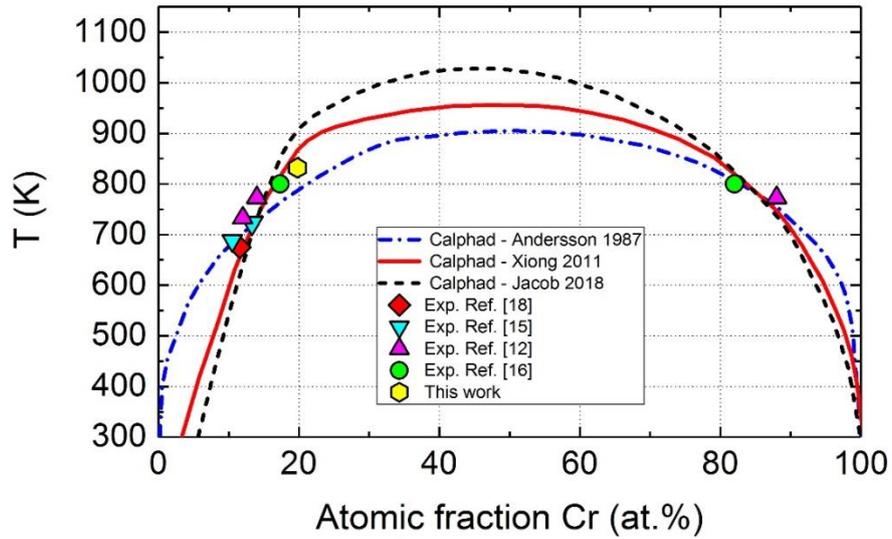

Fig. 6 Comparison between three theoretical Calphad-based predictions concerning the full MG adopted from [10] and our experimental data obtained with the Mössbauer spectroscopy. The size of the corresponding symbols is comparable with the experimental error.

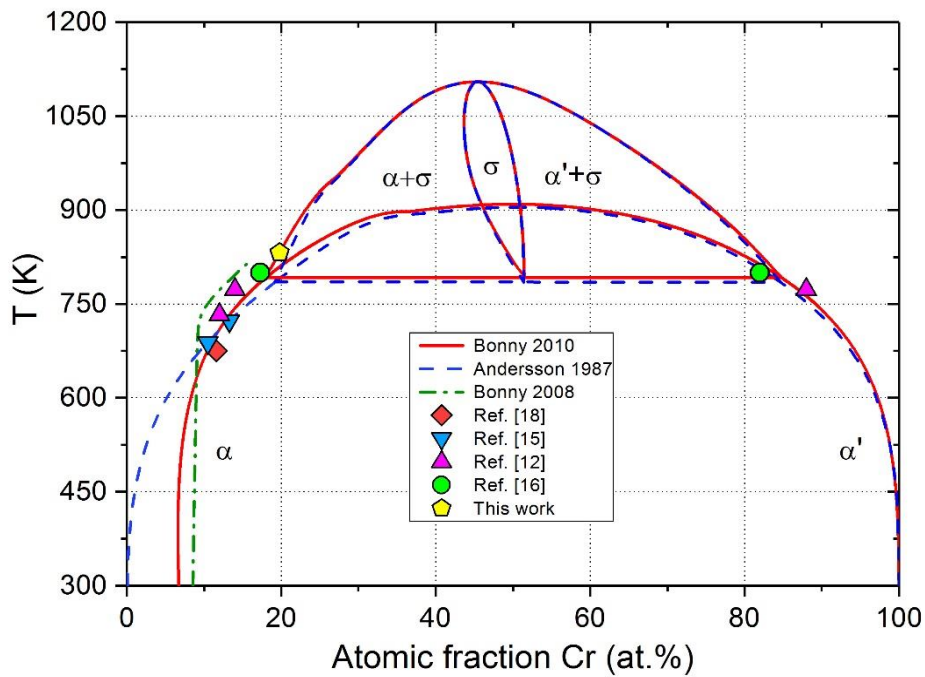

Fig. 7 Relation between experimental data obtained with the Mössbauer spectroscopy and the calculated phase diagram of the Fe-Cr system adopted from [4]. Size of the symbols corresponds to experimental uncertainty.

### 3.2. (0,0) Configuration

Another characteristic feature of the p(H)-curves that signifies the phase decomposition and resulting clustering of Cr atoms is an increase of the intensity of the peak situated at ~335 kOe. This peak can be related to those Fe atoms that have no Cr atoms in their 1NN-2NN neighborhood (in our notation (0,0) atomic configuration). As displayed in Table 1 and visualized in Fig. 8, the probability of this configuration, P(0,0), changes with the annealing time, $t$, in a way similar to that observed in <H> - see Fig. 2. Therefore, the data were analyzed in terms of the JMAK-like equation in two stages separately. The best-fit kinetic parameters are displayed in Fig. 8. They differ from those obtained from the <H>-t dependence what is rather expected as P(0,0) represents only one from 17 atomic configurations taken into account while <H> gives information over all of them, hence the latter is a better indicator of what occurs under annealing.

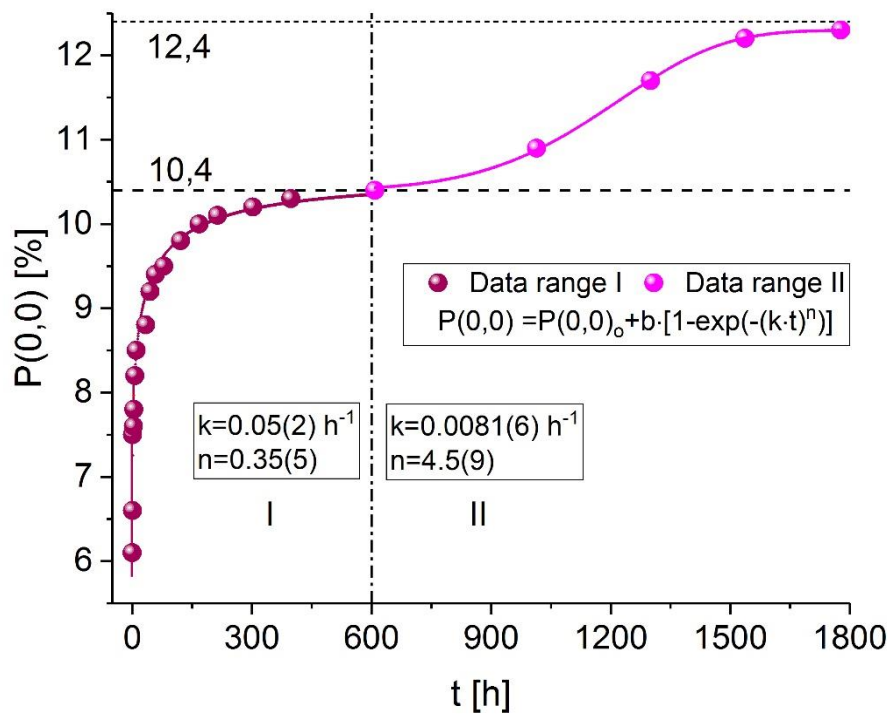

Fig. 8 Probability of the (0,0) atomic configuration, P(0,0), vs. annealing time, *t*. The solid lines stand for the best-fits to the data in terms of the JMAK equation. Values of the rate constant, *k*, and the ones of the Avrami exponent, *n*, are displayed in the legends. Values of P(0,0) in saturation for both ranges are displayed.

The P(0,0)-value of the untreated sample is equal to 6.1 %, whereas the corresponding value expected for the random distribution of Cr atoms is 1.4 %. This shows that the original distribution of Cr atoms in the studied sample was not random. Indeed, if one uses the value of 6.1 % to estimate the concentration of Cr in a random alloy, one would arrive at $x = 1 - \sqrt[14]{0.061}$ =18 at.%, hence much less than the value determined from the chemical analysis (26.3 at.%). This can be interpreted in favor of clustering of Cr atoms in the untreated sample. The maximum value of P(0,0) equals to 12.2 % which corresponds to *x*=14 at.% Cr in a Fe-Cr alloy with the random distribution of atoms. In other words, from the view point of an observer situated at Fe atoms that have no Cr neighbors within the 1NN-2NN neighborhood the concentration of Cr decreased after 1777 h of annealing by 4 at.% relative to its initial value.

### 3.3. Short-range order

The annealing time behaviors of <H> and that of P(0,0) give evidence on clustering of Cr atoms. The effect can be quantitatively expressed in terms of short-range order (SRO) parameters, $\alpha_k$. They can be defined as follows:

$$\alpha_1 = 1 - \frac{\langle m \rangle}{\langle m_r \rangle} \qquad (5a)$$

$$\alpha_2 = 1 - \frac{\langle n \rangle}{\langle n_r \rangle} \qquad (5b)$$

Where $<m> = \sum_{m,n} mP(m,n)$ is the average number of Cr atoms in 1NN; $<n> = \sum_{m,n} nP(m,n)$ is the average number of Cr atoms in 2NN. Corresponding symbols with the subscript *r* stand for the values expected for the random distribution, hence $<m_r> = 8x, <n_r> = 6x$.

The analysis of the spectra in terms of the applied superposition method permitted determination of values of <m>, <n>, hence via equations (5a)-(5c), values of $\alpha_k$ (k=1,2). Figure 9 illustrates the annealing time dependences of $\alpha_1$ and $\alpha_2$. Both of them are positive, what means that in both neighbor shells the number of Cr atoms is lower than expected for the random distribution. This indicates that the effective interaction potential between Fe and Cr atoms is repulsive. However, the two SRO parameters show a different behavior as a function of $t$. While the behavior of $\alpha_1$ resembles that of <H> i.e. it exhibits two stages with a border at ~300 h, $\alpha_2$ initially follows $\alpha_1$ but rather quickly (at ~80 h) achieves its maximum followed by an exponential-like decrease up to ~1000 h. For longer annealing times a weak increase of $\alpha_2$ can be seen. It happens in the time interval where precipitation of $\sigma$ was revealed. The time behavior of $\alpha_1$ has been analyzed using the JMAK equation. The best-fit lines to the data are presented in Fig.9 as solid lines and the best-fit kinetic parameters obtained are displayed in Fig. 10 and also in Table 2.

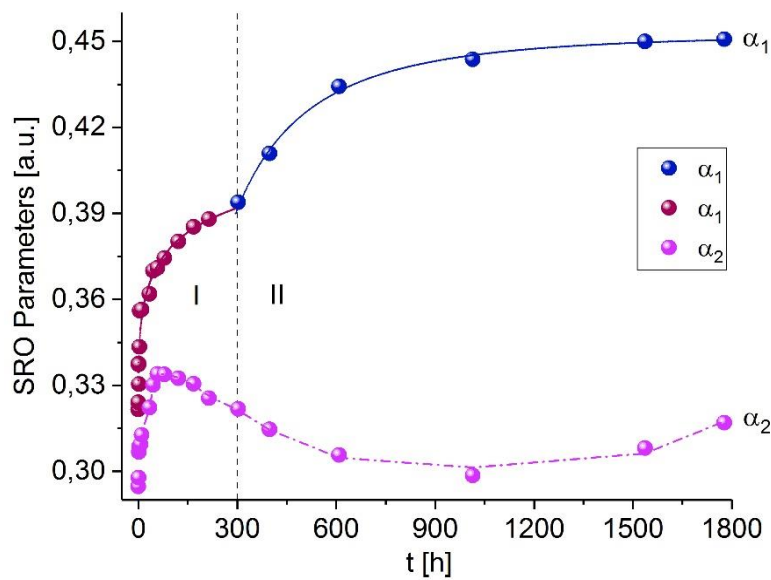

Fig. 9 SRO parameters $\alpha_1$ and $\alpha_2$ vs. annealing time, $t$. Solid lines stand for the best-fit of $\alpha_1$ data to the JMAK equation in the two ranges divided by a vertical dash line. The broken line is the guide to the eye.

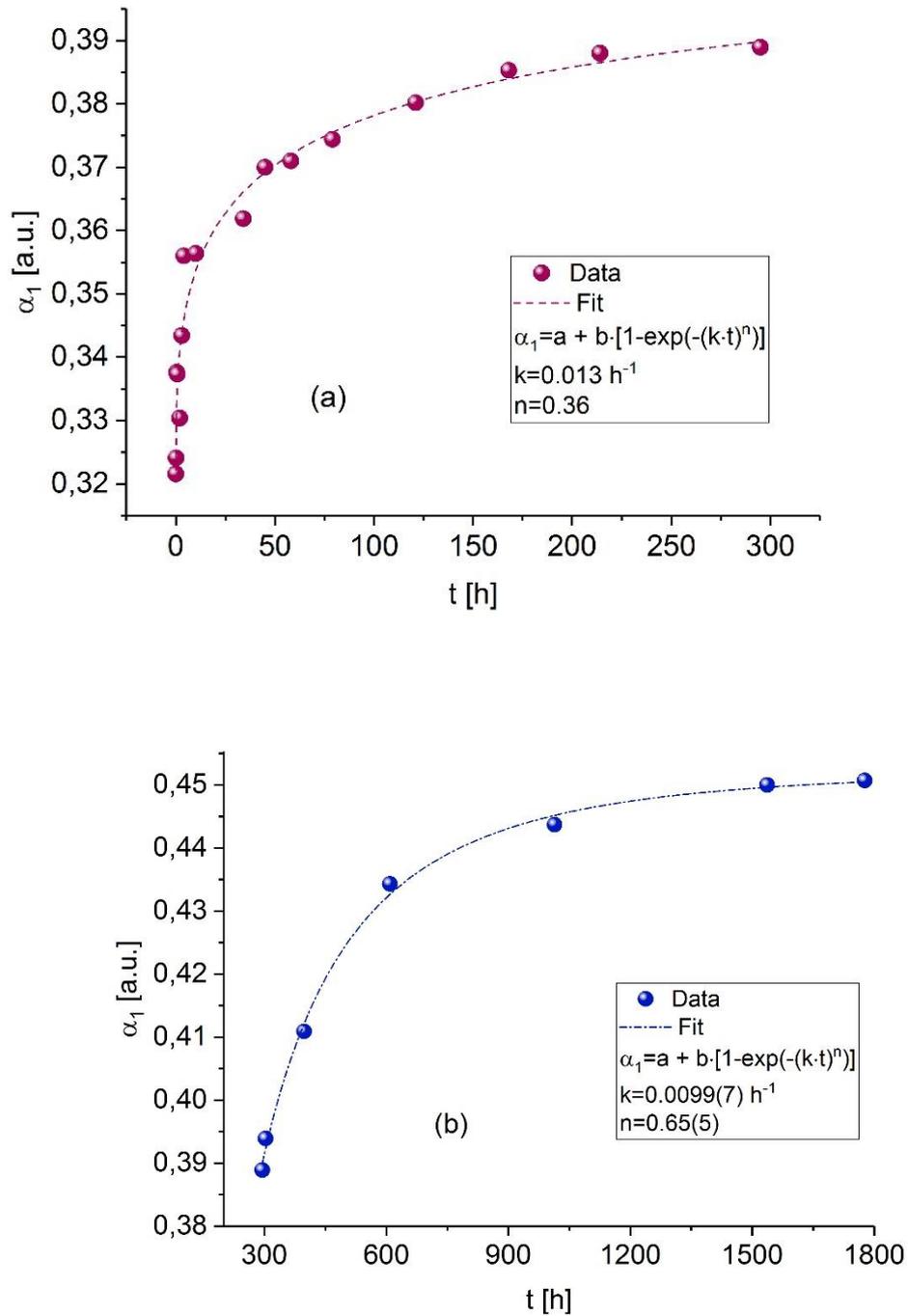

Fig.10 SRO parameter $\alpha_1$ vs. annealing time, *t*, for the range: (a) I and (b) II. The lines are the best-fits of the data to the JMAK equation. Kinetic parameters *k* and *n* are displayed in the legends.

### 3.4. Magnetic texture

Mössbauer spectra are also sensitive to an angle between the direction of the γ-rays (normal to the sample's surface in this experiment), $\Theta$, and the magnetization vector, **M**, hence their analysis can yield value of $\Theta$. In particular $\Theta=90^o$ means that **M** lies in the plane of the sample and $\Theta=0^o$ signifies that **M** is perpendicular to the sample's surface. In the course of annealing of a polycrystalline sample, like in the present case, the orientation of grains can change because of their growth in a preferential direction causing thereby a rotation of **M**. This process can be followed by recording spectra on annealed samples vs. annealing time as was the case in this study. Figure 11 illustrates the $\Theta(t)$ dependence obtained by the applied analysis. It shows a saturation-like behavior and the data could have been nicely fitted to the JMAK equation yielding the kinetics parameters $k$ and $n$ (In the fit we have omitted the value of $\Theta$ for $t$=0 h because it was significantly smaller than the value derived from the spectrum recorded on the sample annealed for 0.12 h. The reason for the large difference was obviously the strain in the untreated sample).

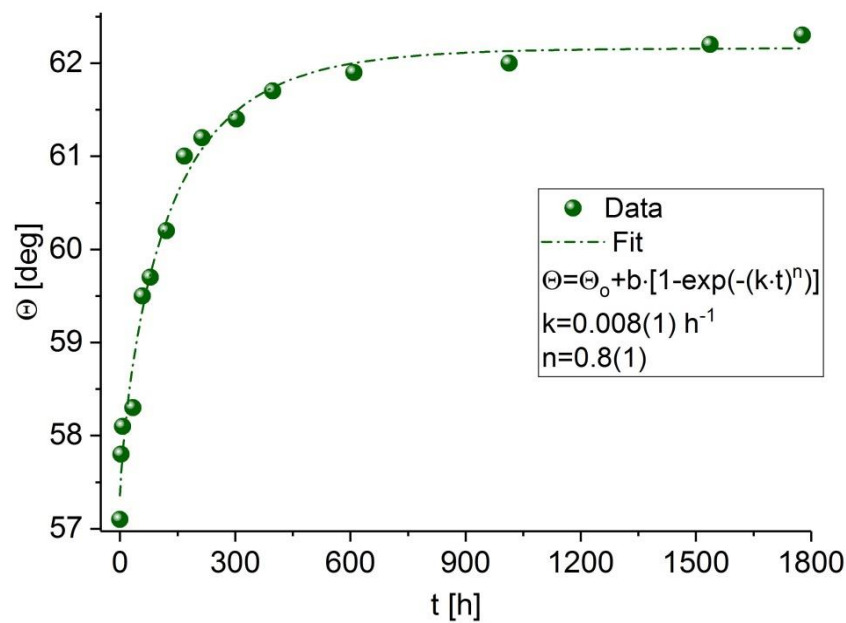

Fig. 11 Annealing time dependence of the average angle between the sample's magnetization vector and the normal to the sample's surface, $\Theta$. The best-fit to the data to the JMAK equation is shown as a dash-dot line. The best-fit kinetic parameters are displayed in the legend.

## 4. Discussion

The present study yielded clear evidence that the $Fe_{73.7}Cr_{26.3}$ alloy isothermally annealed at 832 K decomposes into Fe-rich phase and the σ-phase, and the decomposition process takes place in two stages. Our recent study performed on a sample having the same composition and origin demonstrated that its annealing at 800 K resulted in a phase separation into Fe-rich (α) and Cr-rich (α') phases [16]. It follows from these two experiments that the border line between the phase field where a true miscibility gap exists and the one where the σ-phase can precipitate lies between 800 and 832 K. Noteworthy, this temperature is higher than predicted ones viz. ~785 K [2], ~794 K [4] and ~796 K [10]. The evidence in favor of the two stage decomposition and precipitation processes, as observed in the present study, comes from three quantities viz. the average hyperfine field, <H>, probability of the (0,0) atomic configuration, P(0,0), and the SRO parameter, $α_1$, as illustrated in Figs. 3, 7 and 8, respectively. The dependences of these quantities vs. annealing time are in line with the JMAK. The fits of the corresponding data to this equation yielded kinetics parameters i.e. the Avrami exponent, *n*, and the rate constant, *k*, The value of *n* lies between ~0.4 and ~0.8 what means that the mechanism underlying the phase decomposition and/or precipitation of σ is diffusion-controlled growth and nucleation of isolated platelets and/or needles of finite size, whereas the platelets thicken after their edges have impinged [19]. In turn, the values of *k* give evidence that the first stage process is significantly faster than the one underlying the second stage. The activation energy of the latter has been estimated to be by ~12 kJ/mol higher. The first stage of the decomposition, which takes place in the time span up to ~300 h, can be associated with the decomposition into the Fe-rich phase, which follows on one hand from the increase of the average hyperfine field, <H>, and on the other hand, from the increase of the SRO parameter, $α_1$. Both signify to a decrease of number of Cr atoms in the vicinity of the probe Fe atoms, hence clustering of Cr atoms. The saturation of both quantities indicates a termination of the process. From the <H>-value in saturation the concentration of Cr in the Fe-rich phase was determined as 20.8 at. %. However, on longer annealing <H> and $α_1$ start to increase again what means that the previous saturation was metastable. In other words, the value of 20.8 at.% derived therefrom can be regarded as the border of the metastable miscibility gap. The second stage is much slower, as mentioned above, but it also has a saturation-like character and results in precipitation of 0.6 % of σ. The value of <H> in saturation is here higher and the corresponding concentration of Cr lower viz. 19.8 at.%. The latter can be regarded as a limit of Cr solubility in iron at 832 K. Interestingly, the process of the decomposition has been reflected in a change of the magnetic texture which in the present experiment could be quantitatively followed by an average angle between the magnetization vector and the normal to the sample's surface (or direction of the γ-rays), Θ. As illustrated in Fig. 11, Θ rotates from ~57° to ~62°. This process can be also described in terms of the JMAK equation, however, in this case there is no indication that two stages take place.

## 5. Conclusions

The results obtained in the present study can be concluded as follows:

1. Isothermal annealing of the $Fe_{73.7}Cr_{26.3}$ sample at 832 K resulted in its decomposition into Fe-rich phase and precipitation of the σ-phase.

2. The kinetics of the decomposition process could be quantitatively followed by considering annealing time dependence of several quantities like: average hyperfine field, <H>, short-range order parameter, $\alpha_1$, probability of atomic configuration with no Cr atoms within the first two coordination shells, P(0,0) and angle between the γ-rays and magnetization vector, Θ.

3. The decomposition process took place in two stages clearly visible in the time behavior of <H>, $\alpha_1$ and P(0,0). Both stages could have been well described in terms of the JAMK equation yielding values of the rate constant, *k*, and the Avrami exponent, *n*.

4. The decomposition process in the first stage is significantly faster than the one in the second stage.

5. The value of *n* ranges between ~0.4 and ~0.8 testifying that the process underlying the phase decomposition and/or precipitation of σ is a diffusion-controlled growth and nucleation of isolated platelets and/or needles of finite size combined with their thickening.

6. The border of the metastable miscibility gap has been determined from the <H>-value in saturation in the first stage.

7. The limit of Cr solubility in iron has been determined from the <H>-value in saturation in the second stage.

8. Comparison of the obtained Cr concentration values with theoretical predictions has been made.

9. Average angle between the normal to the sample's surface and the magnetization vector rotates with annealing time by ~$5^o$ in maximum.

**Acknowledgements**

This work was financed by the Faculty of Physics and Applied Computer Science AGH UST and ACMIN AGH UST statutory tasks within subsidy of Ministry of Science and Higher Education. G. Bonny is gratefully thanked for making available his calculations shown in Figs.5 and 7.

Table 2 Values of the kinetic parameters $k$ and $n$ as found by fitting the listed quantities to the JMAK equation, the difference in the energy between the second and the first stage of the decomposition and/or precipitation, $\Delta E$, and the concentration of Cr, $x$, after termination of the first (I) and the second (II) stage of decomposition/precipitation.

| Quantity | $k$ [h$^{-1}$] | $k_I/k_{II}$ | $n$ | $\Delta E$ [kJ/mol] | $x$ [at.%] |
|---|---|---|---|---|---|
| $\langle H \rangle_I$ | 0.022 | 5.5 | 0.64 | 11.6 | 20.7 |
| $\langle H \rangle_{II}$ | 0.004 | | 0.7 | | 19.8 |
| $P(0,0)_I$ | 0.05 | 6.3 | 0.35 | 12.7 | 14.9 |
| $P(0,0)_{II}$ | 0.008 | | 4.5 | | 13.9 |
| $\alpha_1(I)$ | 0.013 | 1.3 | 0.36 | 1.8 | - |
| $\alpha_1(II)$ | 0.010 | | 0.36 | | - |
| $\Theta$ | 0.008 | - | 0.8 | - | - |